\documentclass[aps,prc,twocolumn,showpacs,preprintnumbers,amsmath,amssymb]{revtex4-1}
\usepackage{graphicx}
\usepackage{color}
\definecolor{purple}{rgb}{0.6,0,0.5}
\definecolor{verde}{rgb}{0,0.5,0}
\usepackage[colorlinks=true, pdfstartview=FitV, linkcolor=purple,
citecolor=blue,urlcolor=navyblue]{hyperref}

\begin{document}

\title{Correlations between the nuclear matter symmetry energy, its slope, 
and curvature from a nonrelativistic solvable approach and beyond} 

\author{B. M. Santos$^1$, M. Dutra$^2$, O. Louren\c{c}o$^3$, and A. Delfino$^1$}

\affiliation{$^1$Instituto de F\'isica, Universidade Federal Fluminense,
24210-150, Niter\'oi, RJ, Brazil  \\
$^2$Departamento de F\'isica e Matem\'atica - ICT, 
Universidade Federal Fluminense, 28895-532 Rio das Ostras, RJ, Brazil \\
$^3$Departamento de Ci\^encias da Natureza, Matem\'atica e 
Educa\c c\~ao, CCA, Universidade Federal de S\~ao Carlos, 13600-970 Araras, SP,
Brazil}


\begin{abstract}
By using point-coupling versions of finite range nuclear relativistic
mean field models containing cubic and quartic self interactions in the scalar
field $\sigma$, a nonrelativistic limit is achieved. This approach allows an
analytical expression for the symmetry energy ($J$) as a function of its 
slope ($L$) in a unified form, namely, $\,L\,=\,3J\,+f(m^{*},\rho_{o},B_{o},K_{o})$,
where the quantities $m^{*}$, $\rho_{o}$, $B_{o}$ and $K_{o}$ are bulk parameters at the
nuclear matter saturation density $\rho_{o}$. This result establishes a linear
correlation between $L$ and $J$ which is reinforced by exact relativistic calculations.
An analogous analytical correlation is also found for $J$, $L$ and the symmetry energy
curvature ($K_{\mbox{\tiny sym}}$). Based on these results, we propose graphic constraints
in $L\times J$ and $K_{\mbox{\tiny sym}}\times L$ planes which finite range models must
satisfy. 
\end{abstract}

\maketitle  

\section{Introduction} 

Several bulk parameter quantities help understanding the nuclear matter properties. One of
them is the symmetry energy ${\cal S}$, that can be expanded as a function
of the nuclear density $\rho$ as ${\cal S} (\rho) = J + Lx + \frac{1}{2}K_{\mbox{\tiny
sym}}x^2 + \frac{1}{6}Q_{\mbox{\tiny sym}}x^3 + \mathcal{O}(x^{4})$, where $x\,=\, (\rho
-\rho_{o})/3\rho_{o}$ and $\rho_{o}$ is the nuclear  matter saturation density. 
The coefficients of this expansion, namely, $J$, $L$, $K_{\mbox{\tiny sym}}$ and
$Q_{\mbox{\tiny sym}}$ are, respectively, the symmetry energy at the saturation
density, the slope, curvature, and third derivative (skewness) of $\cal {S}$, all
of them also evaluated at $\rho=\rho_{o}$. The symmetry energy is important to
model nuclear matter and finite nuclei, by probing the isospin part of nuclear 
interactions. Particularly, it is also important in different issues of 
astrophysics~\cite{chenprl,dutra}. For a study of the effects of $J$ and
$L$ on neutron star properties such as the minimum mass that enables the URCA effect, see,
for instance, Ref.~\cite{deb}.

A compelling feature of the nuclear matter bulk parameter study has been to investigate
correlations among them. The investigation on correlations between observables 
is an important issue in physics since the knowledge of one observable may carry 
information about other. In nuclear physics, particularly, an exact nucleon-nucleon 
interaction is unknown, which leads this area to deal with different proposals of 
nuclear forces. Usually, the free parameters of nuclear models are  eliminated in 
favor of a set of observables. Therefore, in nuclear physics, correlations between 
two observables acquire an enormous importance because it reduces the set of 
independent relevant quantities to be used in the nuclear models construction, 
avoiding redundant free parameters fittings~\cite{PLB634-185}. There are few well 
established correlations between nuclear bulk parameters. One of them, usually known as 
Coester line~\cite{coester}, correlates $\rho_{o}$ and the nuclear matter binding 
energy $B_{o}$. Another one was studied by Furnstahl-Rusnak-Serot
(FRS)~\cite{ls-splitting} and reports the correlation between the finite nuclei spin-orbit
splittings and the ratio $m^*=M^*_o/M$ for a family of effective finite range (FR)
relativistic mean-field (RMF) models. $M^*_o$ is the Dirac effective mass of 
the nucleon in symmetric nuclear matter at $\rho=\rho_o$. The results show that 
good values for these splittings are obtained by a restricted class of FR models 
that present $m^*$ in a range of $0.58\leqslant m^*\leqslant 0.64$. Hereafter we 
will refer this range as the FRS constraint. Recently, a correlation between $L$ and 
$J$ has been verified by Ducoin {\it et~al.}~\cite{ducoin} for a set of effective 
relativistic and nonrelativistic nuclear models. Such a study was based on numerical 
results for $J$ and $L$, obtained from different parametrizations. 
We also call the reader attention for previous investigations on analytical 
expressions for $J$ and $L$ in relativistic and nonrelativistic many-nucleon models 
in Refs.~\cite{PRC85,PLB711}. 
 
Theoretically, $J$ and $L$ are expected to be 
constrained~~\cite{PNP58-587,PRC82-024321}. Nevertheless, no analytical relationship 
between these quantities is known up to now. That is why we find important to have a way 
to relate analytically both quantities. In order to proceed in this direction in our 
paper, we have chosen to follow three steps to simplify the FR models which parametrize 
the infinite nuclear matter bulk parameters and finite nuclei 
properties~\cite{fama1,194959,dadi}. First, we select FR models containing
cubic and quartic interactions in the scalar field $\sigma$, i. e., we choose models with
$\sigma^{3}$ and $\sigma^{4}$ contributions in their Lagrangian density. Basically, they
are known as Boguta-Bodmer models~\cite{boguta}. Second, we use their point-coupling
versions~\cite{nlpc2,nlpc3,nlpc4,newrefpc1,newrefpc2,nlpc1}. It is needed to 	
emphasize here that the point-coupling models are as good as the FR ones in the 
description of nuclear matter and finite nuclei. For instance, in Ref.~\cite{nlpc1}, 
the authors were able to obtain, by using a relativistic zero range model, ground state 
binding energies, spin-orbit splittings, and rms charge radii of a large set of closed 
shell nuclei, as well as, of nuclei outside the valley of beta stability (see their Tables 
VIII and X), clearly showing the success of these kind of model. As a side remark, the 
linear point-coupling model and the Walecka one are exactly the same, as one can 
see in Ref.~\cite{bjp}. Third, we perform a nonrelativistic (NR) limit of the 
point-coupling models, based on normalized spinor wave functions after small component 
reduction, exactly in the same way as developed in Ref.~\cite{reinhard}. Such a procedure 
was already used in Ref.~\cite{odilon-nrprc}, in which very good results were found for 
$\rho\leqslant\rho_{o}$. 

Following these steps, we were able to write, in an analytical way, $L$ and
$K_{\mbox{\tiny sym}}$ as a function of $\rho_o$, $B_o$, $m^*$, and $K_o$
(incompressibility at the saturation density). Our results indicate that both
approaches, namely, the NR limit and the \mbox{FR-RMF} models, suggest a decreasing of $L$
when $m^*$ increases whereas the $L$ dependence on $K_o$ is very weak. Similar behavior is
also found regarding the $m^*$ and $K_{o}$ dependence of $K_{\mbox{\tiny sym}}$. In the
case of symmetry energy slope, we also could predict a linear correlation between $L$ and
$J$, that was also supported by the exact \mbox{FR-RMF} calculations.

In particular cases (models presenting close values for $K_o$), our NR calculations 
also indicate another linear correlation for two distinct cases, namely, 
(i) between $K_{\mbox{\tiny sym}}$ and $L$ for fixed values of $J$, or (ii) between 
$K_{\mbox{\tiny sym}}$ and $J$ for fixed values of $L$. These results are also 
confirmed by the relativistic models submitted to the same conditions.

Our paper is organized as follows. In Sec.~\ref{nrnlpc}, we obtain the expressions for
$J$ and $L$ for the NR limit of the point-coupling models, and show how they are
correlated each other. In Sec.~\ref{pred}, we present, based on these correlations,
the predictions on the exact \mbox{FR-RMF} models, also proposing new constraints that
such models should satisfy in order to exhibit good values for finite nuclei spin-orbit
splittings. Finally, in Sec.~\ref{sum-con}, the mainly conclusion are summarized.

\section{The Nonrelativistic limit of nonlinear point-coupling models} 
\label{nrnlpc}

The relativistic nonlinear point-coupling (NLPC) versions of the Boguta-Bodmer
models are described by the following Lagrangian density
\begin{eqnarray}
\mathcal{L}_{\mbox{\tiny NLPC}}&=&\bar{\psi}(i\gamma^{\mu}\partial_{\mu}-M)\psi
-\frac{1}{2}G^{2}_{\mbox{\tiny
V}}(\bar{\psi}\gamma^{\mu}\psi)^{2}+\frac{1}{2}G^{2}_{\mbox{\tiny S}}(\bar{
\psi } \psi)^{2} \nonumber \\
&+&\frac{A}{3}(\bar{\psi}\psi)^{3}+\frac{B}{4}(\bar{\psi}\psi)^{4}-
\frac{1}{2}G^{2}_{\mbox{\tiny TV}}(\bar{\psi}\gamma^{\mu}\vec{\tau}\psi)^{2},
\label{lagrangeana}
\end{eqnarray}
that mimics the two, three and four body point-like 
interactions with the fermionic spinor field $\psi$ associated to the 
nucleon of mass $M$. In this equation, the last term was included 
in order to take into account the asymmetry of the system (different number of
protons and neutrons). In the nonrelativistic limit of the NLPC model, and using the
mean-field approach, the energy density functional at zero temperature for asymmetric
nuclear matter is written as
\begin{eqnarray}
\varepsilon^{\mbox{\tiny (NR)}}(\rho,y)&=&(G^{2}_{\mbox{\tiny V}}-G^{2}_{\mbox{\tiny
S}})\rho^{2} -A\rho^{3} -B\rho^{4} 
\nonumber \\
&+& G^{2}_{\mbox{\tiny TV}}\rho^{2}(2y-1)^{2} +
\frac{3}{10M^*(\rho,y)}\lambda\rho^{\frac{5}{3}},
\label{eassi}
\end{eqnarray}
where the effective mass is
\begin{equation}
M^*(\rho,y)=\frac{M^2}{(M+G^2_{\mbox{\tiny S}}\rho +
2A\rho^2 +3B\rho^3)H_{\frac{5}{3}}},
\end{equation}
with $H_{\frac{5}{3}} =2^{\frac{2}{3}}[y^{\frac{5}{3}}+(1-y)^{\frac{5}{3}}]$,
$\lambda=(3\pi^2/2)^{\frac{2}{3}}$, and \mbox{$y=\rho_p/\rho$} being the proton fraction
of the system. The proton density is $\rho_p$. For a detailed derivation of
Eq.~(\ref{eassi}) from Eq.~(\ref{lagrangeana}) in the $y=1/2$ case, we address the reader
to Ref.~\cite{odilon-nrprc}.

The coupling constants of the model are $G^{2}_{\mbox{\tiny S}}$, 
$G^{2}_{\mbox{\tiny
V}}$, $A$, $B$ and $G^{2}_{\mbox{\tiny TV}}$. The first four of them are adjusted in
order to fix $\rho_o$, $B_o$, $K_o$ and $M^*_o$. This is done by solving a system of four
equations, namely, \mbox{$\varepsilon^{\mbox{\tiny (NR)}}(\rho_o,1/2)=-B_o$},
$K^{\mbox{\tiny
(NR)}}(\rho_o,1/2)=K_o$, $P^{\mbox{\tiny (NR)}}(\rho_o,1/2)=0$ (nuclear matter 
saturation), and $M^*(\rho_o,1/2)=M^*_o$. The pressure and incompressibility are 
defined, respectively, by  $P^{\mbox{\tiny 
(NR)}}(\rho,y)=\rho^2\frac{\partial(\varepsilon^{\mbox{\tiny
(NR)}}/\rho)}{\partial\rho}$ and $K^{\mbox{\tiny (NR)}}(\rho,y)=9\frac{\partial
P^{\mbox{\tiny (NR)}}}{\partial\rho}$.

An advantage of this approach is to obtain simple analytical expressions for the
equations of state (EOS) of the model, in comparison to those calculated in the exact
FR models. It is worth to mention that in the EOS of the NR limit of the NLPC 
models, there are no quantities found in a self-consistent way. All observables are
functions of $\rho$ and $y$, as one can see, for instance, in Eq.~(\ref{eassi}). Thus,
the study of the correlation between the symmetry energy and its slope can be 
performed analytically. For this purpose, we first use Eq.~(\ref{eassi}) to
write $\mathcal{S}(\rho)=\frac{1}{8}\left[\frac{\partial^2(\varepsilon^{\mbox{\tiny
(NR)}}/\rho)}{\partial y^2}\right]_{y=\frac{1}{2}}$. Then, $J=\mathcal{S}(\rho_o)$ is
given by
\begin{eqnarray}
J=\frac{\lambda\rho^{\frac{2}{3}}_o}{6M}
+ \left(G^{2}_{\mbox{\tiny S}} +2A\rho_o +3B\rho_o^{2}\right)
\frac{\lambda\rho^{\frac{5}{3}}_o}{6M^2} + G^2_{\mbox{\tiny TV}}\rho_o.
\label{esymASS-NR}
\end{eqnarray}
The symmetry energy $\mathcal{S(\rho)}$ is used again in order to obtain
$L=3\rho_{o}\left(\frac{\partial {\cal S}}{\partial\rho}\right)_{\rho=\rho_{o}}$. The
result is
\begin{eqnarray}
L=\frac{\lambda\rho_{o}^{\frac{2}{3}}}{3M} 
+ \left(5G^2_{\mbox{\tiny S}} + 16A\rho_o + 33B\rho_{o}^{2}\right)
\frac{\lambda\rho_o^{\frac{5}{3}}}{6M^{2}} + 3G^2_{\mbox{\tiny TV}}\rho_o.\quad
\label{L-NR}
\end{eqnarray}
From Eq.~(\ref{esymASS-NR}) it is possible to determine the last coupling constant
$G^2_{\mbox{\tiny TV}}$, by imposing the model to present a particular value for $J$.

\begin{figure}[!htb]
\centering
\includegraphics[scale=0.3]{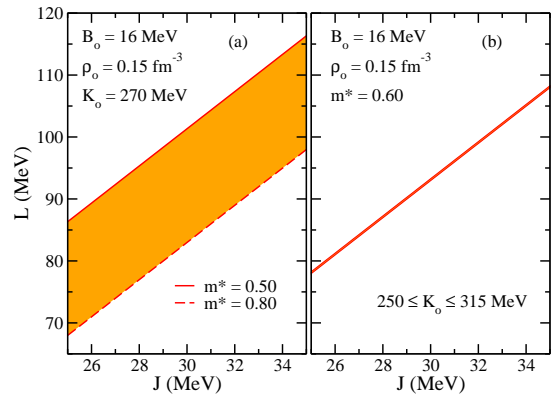}
\vspace{-0.2cm}
\caption{Effect of $\Delta f$ in the $L-J$ correlation of Eq.~(\ref{L-J}) for (a)
$0.50\leqslant m^*\leqslant 0.80$, and (b) $250\leqslant K_o\leqslant 315$~MeV.} 
\label{figdl}
\end{figure}

At this point, we rewrite the coupling constants of the model, namely, 
$G^2_{\mbox{\tiny S}}$, $G^2_{\mbox{\tiny V}}$, $A$, and $B$, in terms of the bulk 
parameters $m^*$, $\rho_o$, $B_o$, and $K_o$. An analogous procedure is done in the 
context of the Skyrme models in Ref.~\cite{sam} through the simulated annealing method. 
Therefore, it is possible to write $L$ explicitly as \mbox{$L=L(m^*,\rho_o, B_o,K_o)$}. 
By doing so, and subtracting $3J$ from $L$, we finally find a clear correlation between 
$J$ and $L$ in the following form
\begin{equation}
L=3J +f(m^{*},\rho_{o},B_{o},K_{o}),
\label{L-J}
\end{equation}
where the function
\begin{eqnarray}
f(m^*,\rho_o,B_o,K_o)=\left(\frac{1}{m^{*}}-1\right)g(\rho_o)+h(\rho_o,B_o,K_o)\quad
\label{functionf}
\end{eqnarray}
exhibits a dependence with the inverse of the effective mass. The functions
$g(\rho_{o})$ and $h(B_{o},K_{o},\rho_{o})$ are written, respectively, as
\begin{align}
g&(\rho_{o})
=\frac{\lambda\rho_o^{\frac{2}{3}}}{3M}
\Bigg[1 + \frac{2E_{\mbox{\tiny F}}^o }{\left(M-2E_{\mbox{\tiny F}}^o \right)}
\nonumber \\
& -\frac{  \left(M-10E_{\mbox{\tiny F}}^o \right)ME_{\mbox{\tiny F}}^o   }
{ \left( 3M^{2} - 19E_{\mbox{\tiny F}}^oM  
+ 18E_{\mbox{\tiny F}}^{o2}\right)\left(M-2E_{\mbox{\tiny F}}^o\right)}\Bigg],
\label{functiong}
\end{align}
and
\begin{align}
h&(\rho_o,B_o,K_o)=-\frac{\lambda\rho_{o}^{\frac{2}{3}}}{6M} \times
\nonumber \\
&\times \Bigg[1 + \frac{2E_{\mbox{\tiny F}}^o\left(M-9E_{\mbox{\tiny F}}^o -27B_{o}
\right) + K_{o}M }{ \left(3M^{2} - 19E_{\mbox{\tiny F}}^oM 
+ 18E_{\mbox{\tiny F}}^{o2}\right)} \Bigg],
\label{functionh}
\end{align}
with $E_{\mbox{\tiny F}}^o=3\lambda\rho_{o}^{\frac{2}{3}}/10M$.
Eqs.~(\ref{L-J})-(\ref{functionh}) contain the main result of our paper. They show, in an
analytical way, a linear correlation between $L$ and $J$. Moreover, if we keep fixed the
values $\rho_o$, $B_o$, and $K_o$, the functions $g(\rho_o)$ and $h(\rho_o,B_o,K_o)$
become constant. Therefore, Eq.~(\ref{L-J}) will exhibit parallel lines for different
$m^*$ values (see Fig.~\ref{figdl}{\color{purple}a}).

Usually, in nuclear mean-field models, the binding energy and the saturation 
density are well established close around the values of \mbox{$B_o=16$~MeV} and 
\mbox{$\rho_o=0.15$~fm$^{-3}$}. The same assumption does not apply to the 
incompressibility and effective mass. Therefore, it is important to see how the 
function $f(m^*,\rho_o,B_o,K_o)$ in Eq.~(\ref{functionf}) varies as a function of 
$K_o$, or $m^*$ for the mentioned values of $B_o$ and $\rho_o$. From 
\mbox{Eqs.~(\ref{functionf})-(\ref{functionh})}, it is straightforward to check that 
for a fixed value of $m^*$, the variation in $f$ will be given by 
\begin{equation}
(\Delta f)_{K_o}=-\frac{\lambda\rho_{o}^{\frac{2}{3}}}{18M^{2} - 114E_{\mbox{\tiny
F}}^oM + 108E_{\mbox{\tiny F}}^{o2}} \Delta K_o.
\label{deltaf}
\end{equation}
For the range of $250\leqslant K_o\leqslant315$~MeV, 
recently proposed in Ref.~\cite{stone}, one can verify that \mbox{$|(\Delta 
f)_{K_o}|=0.32$~MeV}. On the other hand, by choosing two different models presenting 
the same incompressibility $K_{o}$ but with two different effective masses $m_{1}^*$ 
and $m_{2}^* $, the $f$  variation can be inferred by 
\begin{equation}
(\Delta f)_{m^*} =-\frac{g(\rho_o)}{m^*_1m^*_2} \Delta m^*, 
\end{equation}
where $\Delta m^*\,=\,m_{2}^* - m_{1}^*$. For a typical range of \mbox{$0.50\leqslant
m^*\leqslant 0.80$}, presented by \mbox{FR-RMF} models, one has \mbox{$|(\Delta
f)_{m^*}|=18$~MeV}, since \mbox{$g(\rho_o=0.15$~fm$^{-3})=24.5$~MeV}.

Fig.~\ref{figdl} shows how such variations affect the correlation given in 
Eq.~(\ref{L-J}). From this figure we can conclude that different models presenting 
the same effective mass, will exhibit points in a $L\times J$ graph situated very 
close to a same line, since in this case the variation of the linear coefficient in 
Eq.~(\ref{L-J}) is very small compared to that one of the case in which $K_o$ is 
fixed. This leads us to draw the conclusion that in the NR limit of the NLPC models 
described by Eq.~(\ref{lagrangeana}), the linear correlation between $J$ and $L$ in 
Eq.~(\ref{L-J}) is achieved for the more distinct models under the condition that 
their effective masses are equal. Before we end this section, let us remark that 
Refs.~\cite{PRC85,PLB711} could have anticipated a $J\times L$ correlation if the authors 
had worked out their general results for $L(\rho=\rho_o)$ and $E_{\rm sym}(\rho=\rho_o)$. 
Regarding this correlation itself, let us emphasize here that, mathematically, the 
linear behavior is ensured in the NR limit of the NLPC model, since there is only one 
isovector parameter, namely $G^{2}_{\mbox{\tiny TV}}$, in the equations of $J$ and $L$, 
see Eqs.~(\ref{esymASS-NR}) and (\ref{L-NR}). Thus, the result pointed out in 
Eq.~(\ref{L-J}) reflects the limitation of the model parameters, in particular, the 
isovector one. We address to a future work further investigations of possible analytical 
correlations between $J$ and $L$, not necessarily linear, for models with more than one 
isovector parameter.

\section{PREDICTIONS ON \mbox{FR-RMF} MODELS}
\label{pred}

\subsection{SYMMETRY ENERGY SLOPE}

\begin{figure}[!htb]
\centering
\includegraphics[scale=0.3]{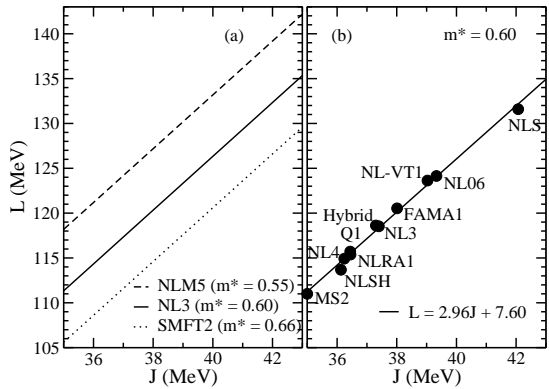}
\vspace{-0.2cm}
\caption{$L$ versus $J$ for (a) the NLM5~\cite{nlm5}, NL3~\cite{nlsh} and 
SMFT2~\cite{smft2} FR parametrizations (see
text) and (b) FR parametrizations in which $m^*$ is the same.} 
\label{rmf-exact}
\end{figure}

Now, we pose the question whether the NR correlation obtained in Eq.~(\ref{L-J}),
and the results showed in Fig.~\ref{figdl} with the subsequently conclusions, still 
remain valid for exact FR models. The answer is given by the study 
we have done for a set of representative FR models, whose results are 
displayed in Fig.~\ref{rmf-exact}.

Fig.~\ref{rmf-exact}{\color{purple}a} shows the $J$ dependence on $L$ for three
different parametrizations of the FR models. For each one of them, we kept fixed 
their respective bulk parameters $m^*$, $\rho_o$, $B_o$, and $K_o$, but allowed 
their symmetry energy $J$ runs. One can verify that for each value of $J$, the 
corresponding $L$, obtained from the relativistic FR models calculations, will be a 
point in a line of angular coefficient equal to $3$. Furthermore, it is also observed 
that $L$ decreases as $m^*$ increases, exactly the same result found in the 
NR limit. In Fig.~\ref{rmf-exact}{\color{purple}b}, we selected a set of FR 
parametrizations~\mbox{\cite{ms2,nlsh,nl4,nlra1,q1,hybrid,fama1,nl06,nl-vt1,
nls}}, presenting the same effective mass, in this case $m^*=0.60$. A best fitting 
curve for these points indicates a line, also pointed out by the NR calculations. 
Moreover, its angular coefficient is given by $2.96$, practically the same number 
found in Eq.~(\ref{L-J}). For a complete list of the \mbox{FR-RMF} models used in 
this work with their mainly saturation properties, we address the reader to the 
Appendix~\ref{appendix}.

As an application of the $J-L$ correlation found in this work, we furnish
a constraint under the values of $L$ for the Boguta-Bodmer FR models. In order to 
do that, we first restrict the range of effective mass to those of the FRS constraint.
Following Ref.~\cite{ls-splitting}, this is the range of $m^*$ that Boguta-Bodmer models
have to be constrained in order to produce spin-orbit splittings in agreement to well 
established experimental values for $^{16}$O, $^{40}$Ca, and $^{208}$Pb. By having this 
constraint as a starting point, we can construct a limiting line defined by $m^*=0.58$ and 
other one at $m^*=0.64$ in a $L\times J$ plane. We have done such lines for the same FR models as in
Fig.~\ref{rmf-exact}{\color{purple}b} by keeping their $\rho_o$, $B_o$ and $K_o$ values,
but changing their effective mass for $m^*=0.58$ and $m^*=0.64$. 
\begin{figure}[!htb]
\centering
\includegraphics[scale=0.3]{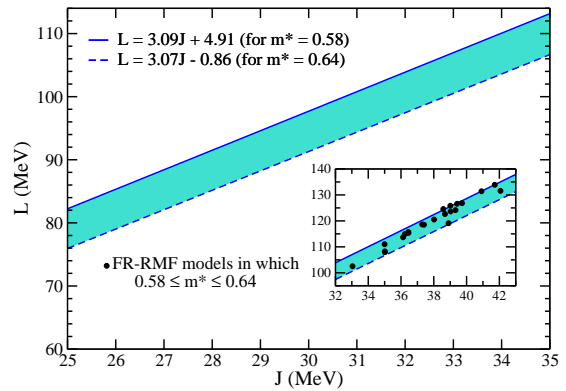}
\vspace{-0.2cm}
\caption{Graphic constraint in the $L\times J$ plane (see text).}
\label{constraint}
\end{figure}
The result is shown in Fig.~\ref{constraint}. Notice that the correlation we have 
found together to the range for the effective mass obtained in 
Ref.~\cite{ls-splitting}, naturally establishes a band of possible values of $L$ as a 
function of $J$ for the Boguta-Bodmer models. In the figure, we show this band in the 
particular range of $25\leqslant J\leqslant 35$~MeV.

In order to test if the \mbox{FR-RMF} models satisfy this constraint, we included
in the inset of Fig.~\ref{constraint} some FR parametrizations
compatible with the FRS constraint, namely, the same of
Fig.~\ref{rmf-exact}{\color{purple}b} together to CS~\cite{cs}, E~\cite{cs}, ER~\cite{cs},
NL3*~\cite{nl3*}, NLB~\cite{nlb}, NLB1~\cite{reinhard}, NLC~\cite{nlb}, NLRA~\cite{nlra},
NLZ~\cite{nl-vt1}, NLZ2~\cite{nl-vt1}, and VT~\cite{cs}. See that all of them fall inside
the band. 

\begin{figure}[!htb]
\centering
\includegraphics[scale=0.3]{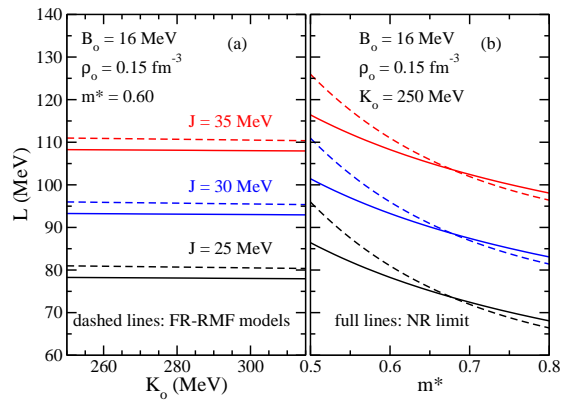}
\vspace{-0.2cm}
\caption{Comparison between the (a) $K_{o}$ and (b) $m^*$ dependences of $L$ for NR
and \mbox{FR-RMF} models at $J=25,30$ and $35$~MeV.}
\label{L1}
\end{figure}

Before we end this subsection we remark here that the NR limit also predicts, for a
fixed value of $J$, correlations between $L$ and the quantities $K_o$ and $m^*$, according
to Eqs.~(\ref{L-J})-(\ref{functionh}). For $m^*$ constant,  $L$ scales as $-K_{o}$ while 
for $K_{o}$ constant, $L$ scales as $1/m^*$. In Fig.~\ref{L1}, we show such dependences
for NR models as well as for the \mbox{FR-RMF} ones.

As we can see, both approaches present the same $L$ variation tendency regarding $K_{o}$
and $m^*$. Notice also that as $m^*$ increases, in the case of $K_o$ fixed
(Fig.~\ref{L1}{\color{purple}b}), the NR limit better approaches the exact \mbox{FR-RMF}
models.

\subsection{SYMMETRY ENERGY CURVATURE} 

In the NR framework it is also possible to find an analytical expression for 
$K_{\mbox{\tiny sym}}=9\rho_o^2\left(\frac{\partial^2 
{\cal S}}{\partial\rho^2}\right)_{\rho=\rho_o}$. It reads
\begin{equation}
K_{\mbox{\tiny sym}}=\left(  \frac{1}{m^*}-1\right)s(\rho_o)
+ r(\rho_o,B_o,K_o)  
\label{k-sym-mass} 
\end{equation}
with
\begin{align}
s(\rho_o)&=\frac{5\lambda\rho_o^{\frac{2}{3}}}{3M}\Bigg[
1+\frac{4E_{\mbox{\tiny F}}^o}{\left(M-2E_{\mbox{\tiny F}}^o \right)} 
\nonumber \\
&- \frac{E_{\mbox{\tiny F}}^o\left(M-10E_{\mbox{\tiny F}}^o 
\right)\left(19M-18E_{\mbox{\tiny F}}^o \right)}
{5 \left(M-2E_{\mbox{\tiny F}}^o \right)\left( 3M^{2} - 19E_{\mbox{\tiny F}}^oM 
+18E_{\mbox{\tiny F}}^{o2} \right)}
\Bigg],
\label{functions}
\end{align}
and
\begin{align}
r(\rho_o,B_o,K_o)&=-\frac{\lambda\rho_o^{\frac{2}{3}}}{3M}\Bigg[ 1 + 
\frac{K_o\left(19M-18E_{\mbox{\tiny F}}^o \right)}{2\left(3M^{2} - 19E_{\mbox{\tiny 
F}}^oM +18E_{\mbox{\tiny F}}^{o2}\right)}
\nonumber \\
&-\frac{(81B_oM + 8 E_{\mbox{\tiny F}}^o M+18E_{\mbox{\tiny F}}^{o2})}{3M^{2} -
19E_{\mbox{\tiny F}}^oM 
+18E_{\mbox{\tiny F}}^{o2}}\Bigg].
\label{functionr}
\end{align} 

By rearranging these equations, we find a simplified form for
$K_{\mbox{\tiny sym}}$, namely,
\begin{equation}
K_{\mbox{\tiny sym}}=\left[  L-3J\right] p(\rho_o) + q(\rho_o,B_o,K_o),
\label{k-sym-L-J} 
\end{equation}
where
\begin{equation}
p(\rho_o)=\frac{s(\rho_o)}{g(\rho_o)},
\label{functionp}
\end{equation}
and
\begin{align}
q&(\rho_o,B_o,K_o)=-h(\rho_o,B_o,K_o)p(\rho_o) +r(\rho_o,B_o,K_o) \nonumber \\
&=\frac{\lambda\rho_{o}^{\frac{2}{3}}}{3M} \Bigg\{\frac{ [p(\rho_o) -2] }{2} 
+ \frac{ ME_{\mbox{\tiny F}}^o [p(\rho_o)+8] }{ \left(3M^{2} - 19E_{\mbox{\tiny F}}^oM 
+ 18E_{\mbox{\tiny F}}^{o2}\right)} \nonumber \\
&- \frac{ 9E_{\mbox{\tiny F}}^{o2} [p(\rho_o)-2] +27B_{o} 
[ E_{\mbox{\tiny F}}^o p(\rho_o) -3M ] }{ \left(3M^{2} - 19E_{\mbox{\tiny F}}^oM 
+ 18E_{\mbox{\tiny F}}^{o2}\right)} \nonumber \\
& + \frac{ M \left[ p(\rho_o) -19 \right]+18E_{\mbox{\tiny F}}^o }{ 2\left(3M^{2} -
19E_{\mbox{\tiny F}}^oM 
+ 18E_{\mbox{\tiny F}}^{o2}\right)}K_{o}  
 \Bigg\}.
\label{functionq}
\end{align} 

Above, $p(\rho_o)=5.13$ at $\rho_o=0.15$~fm$^{-3}$ and a $L$ and $J$ dependence of
$K_{\mbox{\tiny sym}}$ is explicited, see Eq.~(\ref{k-sym-L-J}). {\bf It is worth to 
note that the mathematical relation presented between $K_{\mbox{\tiny sym}}$ and $L$ was 
based on the result of Eq.~(\ref{L-J}), that by itself is a consequence of the limitation 
of the number of isovector parameters of the NR limit of the NLPC model, in this case only 
one, $G^{2}_{\mbox{\tiny TV}}$. For models with two or more isovector parameters, the 
correlation between $J$ and $L$, and consequently, the other between $K_{\mbox{\tiny 
sym}}$ and $L$ (or between $K_{\mbox{\tiny sym}}$ and $J$), may follows a behavior 
different from the linear one.}

Once again, we test whether these results reflect the \mbox{FR-RMF} models calculations. 
Firstly, notice Eq.~(\ref{k-sym-mass}) predicts $K_{\mbox{\tiny sym}}$ constant 
for fixed values of $K_o$ and $m^*$, quite independent of $J$. For a sake of 
illustration we calculate 
$K_{\mbox{\tiny sym}}$ for a set of \mbox{FR-RMF} models presenting
\mbox{$\rho_o=0.15$~fm$^{-3}$}, \mbox{$B_o=16$~MeV}, \mbox{$K_o=270$~MeV},
\mbox{$m^*=0.60$}, and $J$ running in the range of
\mbox{$25\leqslant J\leqslant 35$~MeV}. For these cases, we have obtained a unique value
of \mbox{$K_{\mbox{\tiny sym}}=96.4$~MeV}, supporting the NR prediction of 
Eq. (\ref{k-sym-mass}). 

Still analyzing Eq.~(\ref{k-sym-mass}), we can see that a variation in $m^*$ 
produces a spread in $K_{\mbox{\tiny sym}}$ of 
\begin{equation}
\Delta K_{\mbox{\tiny sym}} =-\frac{s(\rho_o)}{m^*_1m^*_2}\Delta m^*,
\end{equation}
for $K_o$ fixed. Thus, the range of $m^*$ given by the FRS constraint generates
$|\Delta K_{\mbox{\tiny sym}}|=20$~MeV, since \mbox{$s(\rho_o=0.15$~fm$^{-3})=125.7$~MeV}.
A non negligible spread in $K_{\mbox{\tiny sym}}$ is also observed for models with $m^*$
constant and different $K_o$. For such cases, one can see that this spread is entire due
to 
\begin{equation}
\Delta r=-\frac{\lambda\rho_{o}^{\frac{2}{3}}(19M-18E_{\mbox{\tiny F}}^o)}
{18M^3 - 114E_{\mbox{\tiny F}}^oM^2 + 108ME_{\mbox{\tiny F}}^{o2}}\Delta K_o.
\end{equation} 
For the range of \mbox{$250\leqslant K_o\leqslant 315$~MeV}, we calculate $|\Delta
K_{\mbox{\tiny sym}}|=5.9$~MeV. 

\begin{figure}[!htb]
\centering
\includegraphics[scale=0.3]{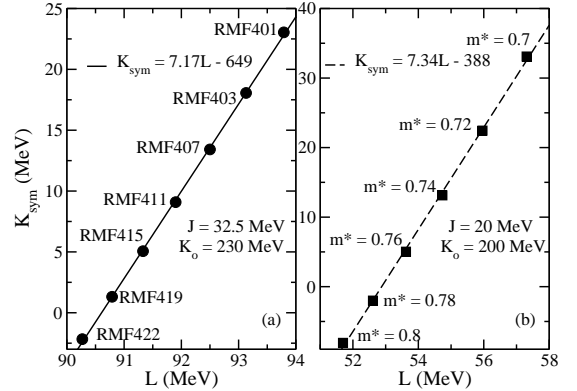}
\vspace{-0.2cm}
\caption{Correlation between $K_{\mbox{\tiny sym}}$ and $L$ (see text). } 
\label{ksym-exact}
\end{figure}

Based on this study and the Eq.~(\ref{k-sym-L-J}), we can conclude, for 
instance, that the linear correlation between $K_{\mbox{\tiny sym}}$ and $L$ for 
$J$ constant will certainly occurs for models in which $\Delta r=0$, i. e., for 
fixed $K_o$. We verified this prediction for the \mbox{FR-RMF} models of 
Ref.~\cite{dadi} with \mbox{$J=32.5$~MeV} and \mbox{$K_o=230$~MeV}. The result is 
depicted in Fig.~\ref{ksym-exact}{\color{purple}a}.

The figure clearly confirms the prediction of Eq.~(\ref{k-sym-L-J}). Even the angular 
coefficient, $7.17$, is comparable with the value $p(\rho_o)=5.13$ (at 
$\rho_o=0.15$~fm$^{-3}$) of the NR calculation.

We remark here that the correlation observed in Fig.~\ref{ksym-exact}{\color{purple}a} 
only occurs for models with the same values for $K_o$ and $J$ ($230$~MeV and $32.5$~MeV, 
respectively, in this case). If $J$ is not the same, $K_{\mbox{\tiny sym}}$ will be random 
and independent of the value of $L$. Indeed, most FR-RMF models in the literature can 
suggest that $K_{\mbox{\tiny sym}}$ first decreases with the increase of $L$ and attains 
a minimum at about $L=70$~MeV, then rises back for larger $L$. We reinforce that 
our study indicates that in the Boguta-Bodmer models, such analysis must take into 
account the values of $K_o$ and $J$ of the parametrization, in the sense that only 
with these values fixed, the linear correlation between $K_{\mbox{\tiny sym}}$ and 
$L$ will be established. In order to show that $K_{\mbox{\tiny sym}}\propto L$ even 
for parametrizations presenting $L<70$~MeV, we have constructed Boguta-Bodmer 
models with fixed bulk parameter $\rho_o=0.15$~fm$^{-3}$, $B_o=16$~MeV, $K_o=200$~MeV, 
$J=20$~MeV, and with $m^*$ in the range of \mbox{$0.7\leqslant m^* \leqslant 0.8$}. In 
such case, the parameterizations present $L<70$~MeV, and one can see from 
Fig.~\ref{ksym-exact}{\color{purple}b} that the linear correlation between $K_{\mbox{\tiny 
sym}}$ and $L$ is preserved. Notice, however, that as the value $J=20$~MeV is 
actually ruled out by experimental evidences, our analysis suggests that despite 
mathematically valid for $L<70$~MeV, the correlation between $K_{\mbox{\tiny sym}}$ and 
$L$ for Boguta-Bodmer models, predicts higher values for the symmetry energy slope, see 
Fig.~\ref{ksym-exact}{\color{purple}a}. This is a direct consequence of the model 
structure itself, regarding the number of free isovector parameters. Indeed, the 
prediction of higher $L$ values for acceptable $J$ values can also be seen in 
Fig.~\ref{constraint}.

\begin{figure}[!htb]
\centering
\includegraphics[scale=0.3]{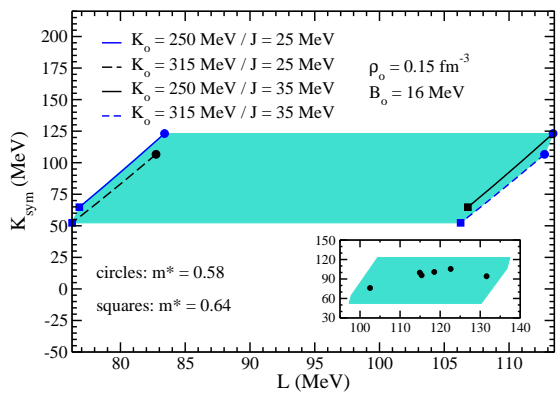}
\vspace{-0.2cm}
\caption{Graphic constraint in the $K_{\mbox{\tiny sym}}\times L$ plane (see text).}
\label{ksyml}
\end{figure}

For the sake of completeness, we use the correlation between 
the symmetry energy, its slope and curvature to propose another graphic 
constraint in the $K_{\mbox{\tiny sym}}\times L$ plane that \mbox{FR-RMF} models 
must satisfy. For this purpose, we used the relativistic framework to 
construct boundaries in that plane by observing the FRS constraint 
and the ranges of $250\leqslant K_o\leqslant315$~MeV and $25\leqslant
J\leqslant 35$~MeV. In such boundaries, we have fixed the values of
$\rho_o=0.15$~fm$^{-3}$ and $B_o=16$~MeV. This procedure leads to the 
band showed in Fig.~\ref{ksyml}, i. e., all \mbox{FR-RMF} models presenting 
$m^*$, $K_o$ and $J$ in the mentioned ranges must produce points in 
the $K_{\mbox{\tiny sym}}\times L$ graph inside this band. In order to 
test this prediction, we selected the same \mbox{\mbox{FR-RMF}}
parametrizations of the inset of Fig.~\ref{constraint} presenting 
$250 \leqslant K_o \leqslant 315$~MeV and reconstructed the band 
in the $K_{\mbox{\tiny sym}}\times L$ plane to take into account that 
such models have $32\leqslant J\leqslant 43$~MeV (see inset of
Fig.~\ref{constraint}). The new band is represented in the inset 
of Fig.~\ref{ksyml}. See that all \mbox{\mbox{FR-RMF}} models (NL3*, NLS, 
NL4, NL3, NLB1, and NLRA1), represented by the
full circles, fall inside the band. In the next section we discuss 
this correlation in a more critical way. 

\begin{figure}[!htb]
\centering
\includegraphics[scale=0.3]{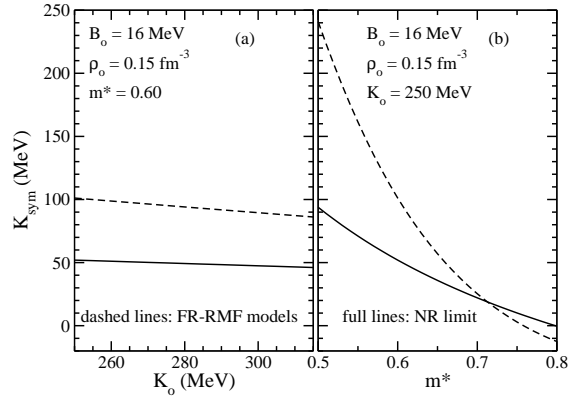}
\vspace{-0.2cm}
\caption{Comparison between the (a) $K_{o}$ and (b) $m^*$ dependences of
$K_{\mbox{\tiny sym}}$ for NR and \mbox{FR-RMF} models.}
\label{K1}
\end{figure}

As a last remark, we proceed to show the $K_{o}$ and $m^*$ dependence of
$K_{\mbox{\tiny sym}}$ by using Eq.~(\ref{k-sym-mass}). Notice that, exactly as in the
case of the symmetry energy slope, Eq.~(\ref{k-sym-mass}) express clear correlations
of $K_{\mbox{\tiny sym}}$ with the incompressibility and effective mass, namely,
$K_{\mbox{\tiny sym}}\sim -K_o$ and $K_{\mbox{\tiny sym}}\sim 1/m^*$ for fixed values of
$m^*$ or $K_o$, respectively. This behavior is depicted in Fig.~\ref{K1}, which also shows
a direct comparison between the results for the NR and \mbox{FR-RMF} approaches.
               
As in the case of the symmetry energy slope, we see that as $m^*$ increases,
$K_{\mbox{\tiny sym}}$ decreases, while for both NR and \mbox{FR-RMF}, the $K_o$
dependence of $K_{\mbox{\tiny sym}}$ is very weak compared to the $m^*$ one. 

\section{Summary and conclusion} 
\label{sum-con}
Since in the majority of the RMF models, $B_{o}$ and $\rho_{o}$ are chosen to be very
close to $16$~MeV and $0.15$ fm$^{-3}$, respectively, in this section, we will forget such
model dependence on them. Here, we are committed with the $m^*$ and the $K_{o}$
dependences. 

In summary, the study performed in this paper indicates that the nonrelativistic limit of
the NLPC models described by Eq.~(\ref{lagrangeana}), can be used as a suitable guideline
to infer possible correlations related to the FR relativistic models with $\sigma^3$ and
$\sigma^4$ self interactions. Regarding the correlations between the quantities at
the saturation density ($\rho=\rho_{o}$), obtained from the nonrelativistic limit and
reproduced by the FR relativistic models, our main findings are the following:

$\bullet$ In the NR approximation, the symmetry energy slope $L$ is linearly
correlated with $J$, see Eq.~(\ref{L-J}). Moreover, this same equation shows that $L$ also
depends explicitly on the effective nucleon mass $m^*$ (scales as $1/m^*$, see
\mbox{Eqs.~(\ref{L-J})-(\ref{functionf})}) and the incompressibility $K_o$ (scales
linearly, see \mbox{Eqs.~(\ref{L-J})-(\ref{functionh})}). The $K_o$ dependence of
$L$ has been verified to be negligible as shown by the full lines of
Fig.~\ref{L1}{\color{purple}a}. We verified that the same features are also found in the
\mbox{FR-RMF} Boguta-Bodmer models, as one can see in Fig.~\ref{rmf-exact} and in the
dashed lines of Fig.~\ref{L1}.

$\bullet$ The symmetry energy curvature $K_{\mbox{\tiny sym}}$ depends on $m^*$,
scaling as $1/m^{*}$, and is linearly correlated with $K_{o}$ in the NR approach, see
\mbox{Eqs.~(\ref{k-sym-mass})-(\ref{functionr})}. Such dependences are not negligible.
By aiming to find a $L$ (or $J$) dependence in $K_{\mbox{\tiny sym}}$, we have
rewritten $K_{\mbox{\tiny sym}}$ as presented in Eq.~(\ref{k-sym-L-J}),
\mbox{$K_{\mbox{\tiny sym}}=\left[L-3J\right] p(\rho_o) + q(\rho_o,B_o,K_o)$}. However,
the existing correlation between $L$ and $J$, see Eq.~(\ref{L-J}), shows that for a fixed 
value of $K_o$, there are two possible scenarios, namely, (i) a linear correlation between 
$K_{\mbox{\tiny sym}}$ and $L$ for models in which $J$ is the same, and (ii) a linear 
correlation between $K_{\mbox{\tiny sym}}$ and $J$ for models in which $L$ is the same. 
Once again, the same correlations also apply to the \mbox{FR-RMF} Boguta-Bodmer models, as 
displayed in Figs.~\ref{K1} and \ref{ksym-exact}.    

$\bullet$ Convinced of the correlation between $L$ and $J$, found in the NR
approximation and confirmed for the relativistic calculations, we have constructed a
region of possible $L$ values as a function of $J$, and according the FRS
constraint~\cite{ls-splitting}, that \mbox{\mbox{FR-RMF}} Boguta-Bodmer models must
satisfy in order to give values for the finite nuclei spin-orbit splitting compatible with well
established experimental values, see Fig.~\ref{constraint}. 

\begin{figure}[!htb]
\centering
\includegraphics[scale=0.41]{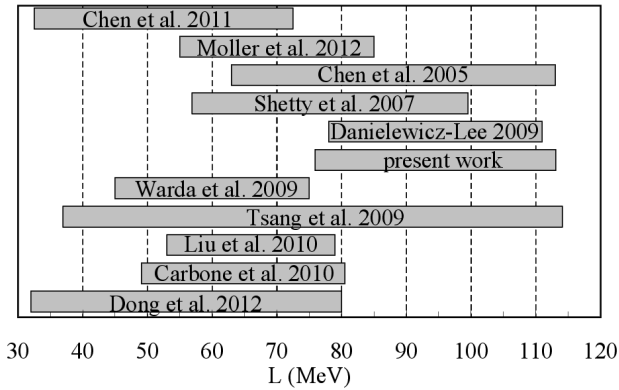}
\vspace{-0.2cm}
\caption{Comparison between the limits of $L$ obtained in this work and those from
Dong {\it et al.}~\cite{PRC85-034308}, Carbone {\it et al.}~\cite{PRC81-041301R}, 
Liu {\it et al.}~\cite{PRC82-064306}, Tsang {\it et al.}~\cite{PRL102-122701}, Warda {\it 
et al.}~\cite{PRC80-024316}, Danielewicz and Lee~\cite{NPA818-36}, Shetty {\it et 
al.}~\cite{PRC75-034602}, Chen {\it et al.}~\cite{PRC72-064309}, M\"oller {\it et 
al.}~\cite{PRL108-052501}, and Chen~\cite{PRC83-044308}.}
\label{lvalues1}
\end{figure}

$\bullet$ In Fig.~\ref{lvalues1}, we present our prediction for the lowest and 
highest values for $L$ in comparison with other values found in the literature, by 
taking into account the region of Fig.~\ref{constraint} in a range of 
\mbox{$25\leqslant J\leqslant 35$~MeV} for the symmetry energy. Notice that our limits for 
$L$ are comparable with other models.

\begin{figure}[!htb]
\centering
\includegraphics[scale=0.35]{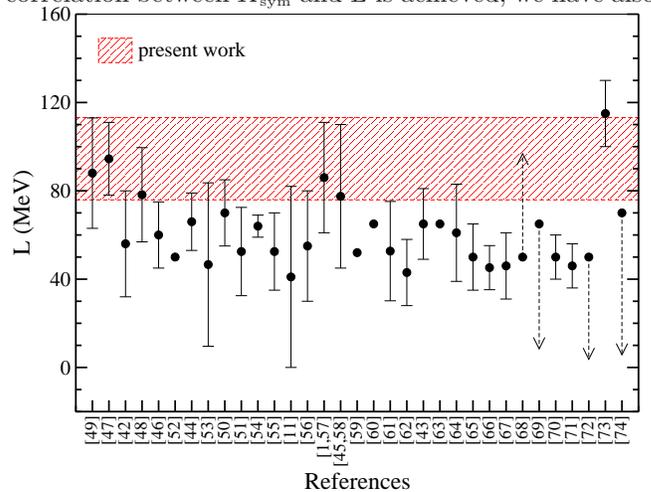}
\vspace{-0.2cm}
\caption{Comparison between the limits of $L$ obtained in this work and those 
from $33$ different analyses of \mbox{Refs.~
\cite{chenprl,PRC82-024321,PRC72-064309,NPA818-36,PRC85-034308,PRC75-034602, 
PRC80-024316,NPA601-141,PRC82-064306,APJ771-51,PRL108-052501,PRC83-044308,
PRL109-262501,NPA992-1,PRL102-122502,PRC72-064611,PRL92-062701,PRL102-122701,
PRC82-051603,PRC76-024606,PRC82-054607,PRC76-051603R,PRC81-041301R,PRC82-064601,
PRC87-014303,PRC88-014302,PLB726-234,arXiv13072706,PRC85-045808,PRC85-025801,
APJ722-33,PRL108-081102,MNRAS418-2343,MNRAS428-L21,PRC80-065809}}.}
\label{lvalues2}
\end{figure}

For the sake of completeness, we present in Fig.~\ref{lvalues2} a large set of $L$ 
values obtained from analyses of different terrestrial nuclear experiments and 
astrophysical observations. They include analysis of isospin diffusion, neutron skin, 
pygmy dipole resonances, $\alpha$ and $\beta$ decays, transverse flow, mass-radius 
relation of neutron stars, torsional crust oscillation of neutron stars, and other more. 
$28$ of the $33$ points showed in the figure were extracted from Table~I of 
Ref.~\cite{PLB727-276}, in which the authors, through the Hugenholtz-Van Hove theorem, 
used these values in order to constraint the neutron-proton effective mass splitting in 
nonrelativistic nuclear models.

$\bullet$ Analogously, but based in the situation in which the correlation
between $K_{\mbox{\tiny sym}}$ and $L$ is achieved, we have also proposed a graphic
constraint in the $K_{\mbox{\tiny sym}}\times L$ plane, that the \mbox{FR-RMF} models at
the FRS condition and presenting \mbox{$25\leqslant J\leqslant 35$~MeV} must obey, see
Fig.~\ref{ksyml}.

Before we end this work, some words of caution are needed. First, we have studied a
particular class of  $\sigma^3\,+\,\sigma^4$ self interactions RMF models where the
nonrelativistic limit used a point-coupling approximation of them. Nevertheless, what we
have called exact calculations in this work has no other approximation than that of the
mean-field, and the point-coupling version of them are absent. Second, our caution words
here are more in the sense that nowadays there are several families of RMF models 
(for a review, see, for instance, Ref.~\cite{dutrarmf}), density dependent among them,
which not necessarily will follow the same features of the \mbox{FR-RMF}
models studied here.

\section*{Acknowledgements}

We thank the partial support from Funda\c c\~ao de Amparo \`a Pesquisa do 
Estado de S\~ao Paulo (FAPESP) and Coordena\c c\~ao de Aperfei\c coamento de 
Pessoal de N\'ivel Superior (CAPES) of Brazil.

\appendix
\section{Saturation properties of the \mbox{FR-RMF models}}
\label{appendix}

In this appendix we show in the following Table~\ref{tab}, the mainly saturation
properties, calculated at the saturation density, of the \mbox{FR-RMF} used in our work.
\onecolumngrid
\hspace{0.5cm}
\begin{table}[!htb]
\begin{ruledtabular}
\caption{Nuclear matter properties, at the saturation density, of the \mbox{FR-RMF} models
used in this work.}
\centering
\begin{tabular}{lcccccrr}
Model & $\rho_o$ & $B_o$ & $K_o$ & $m^*$ & $J$ & $L$~~~ & $K_{\mbox{\tiny sym}}$~~\\
& (fm$^{-3}$) & (MeV) & (MeV) &  & (MeV) & (MeV) & (MeV)\\ 
\hline
CS     & $0.150$ & $16.17$ & $187.21$ & $0.58$ & $40.91$ & $131.42$ & $136.68$ \\ 
E      & $0.150$ & $16.13$ & $221.43$ & $0.58$ & $38.58$ & $124.57$ & $132.12$ \\ 
ER     & $0.149$ & $16.16$ & $220.49$ & $0.58$ & $39.42$ & $126.60$ & $127.62$ \\ 
FAMA1  & $0.148$ & $16.00$ & $200.05$ & $0.60$ & $38.01$ & $120.53$ & $113.22$ \\ 
Hybrid & $0.148$ & $16.24$ & $230.01$ & $0.60$ & $37.30$ & $118.62$ & $110.94$ \\ 
MS2    & $0.148$ & $15.75$ & $249.92$ & $0.60$ & $35.00$ & $111.00$ & $100.85$ \\ 
NL-VT1 & $0.150$ & $16.09$ & $179.03$ & $0.60$ & $39.03$ & $123.63$ & $117.72$ \\ 
NL06   & $0.147$ & $16.05$ & $195.09$ & $0.60$ & $39.33$ & $124.14$ & $110.85$ \\ 
NL3    & $0.148$ & $16.24$ & $271.53$ & $0.60$ & $37.40$ & $118.53$ & $100.88$ \\ 
NL3*   & $0.150$ & $16.31$ & $258.25$ & $0.59$ & $38.68$ & $122.63$ & $105.56$ \\ 
NL4    & $0.148$ & $16.16$ & $270.34$ & $0.60$ & $36.24$ & $114.92$ & $99.72$ \\ 
NLB    & $0.148$ & $15.77$ & $421.02$ & $0.61$ & $35.01$ & $108.26$ & $54.94$ \\ 
NLB1   & $0.162$ & $15.79$ & $280.44$ & $0.62$ & $33.04$ & $102.51$ & $76.15$ \\ 
NLC    & $0.148$ & $15.77$ & $224.46$ & $0.63$ & $35.02$ & $107.97$ & $76.91$ \\ 
NLM5   & $0.160$ & $16.00$ & $200.00$ & $0.55$ & $30.00$ & $103.18$ & $179.44$ \\
NLRA   & $0.157$ & $16.25$ & $320.48$ & $0.63$ & $38.90$ & $119.09$ & $62.11$ \\ 
NLRA1  & $0.147$ & $16.15$ & $285.23$ & $0.60$ & $36.45$ & $115.38$ & $95.72$ \\ 
NLS    & $0.150$ & $16.44$ & $262.94$ & $0.60$ & $42.07$ & $131.59$ & $94.22$ \\ 
NLSH   & $0.146$ & $16.36$ & $355.65$ & $0.60$ & $36.13$ & $113.68$ & $79.83$ \\ 
NLZ    & $0.151$ & $16.18$ & $172.84$ & $0.58$ & $41.72$ & $133.91$ & $140.19$ \\ 
NLZ2   & $0.151$ & $16.06$ & $172.23$ & $0.58$ & $39.01$ & $125.82$ & $140.62$ \\ 
Q1     & $0.148$ & $16.10$ & $241.86$ & $0.60$ & $36.44$ & $115.71$ & $105.65$ \\ 
RMF401 & $0.153$ & $16.30$ & $229.99$ & $0.71$ & $32.50$ & $93.79$ & $23.04$ \\ 
RMF403 & $0.153$ & $16.30$ & $229.99$ & $0.72$ & $32.50$ & $93.13$ & $18.06$ \\ 
RMF407 & $0.153$ & $16.30$ & $229.99$ & $0.73$ & $32.50$ & $92.50$ & $13.42$ \\ 
RMF411 & $0.153$ & $16.30$ & $229.99$ & $0.74$ & $32.50$ & $91.90$ & $9.09$ \\ 
RMF415 & $0.153$ & $16.30$ & $229.98$ & $0.75$ & $32.50$ & $91.33$ & $5.06$ \\ 
RMF419 & $0.153$ & $16.30$ & $229.99$ & $0.76$ & $32.50$ & $90.79$ & $1.31$ \\ 
RMF422 & $0.153$ & $16.30$ & $229.99$ & $0.77$ & $32.50$ & $90.27$ & $-2.17$ \\ 
SMFT2  & $0.162$ & $13.78$ & $211.31$ & $0.66$ & $17.38$ & $52.73$ & $60.27$ \\
VT     & $0.153$ & $16.09$ & $172.74$ & $0.59$ & $39.72$ & $126.83$ & $130.05$ \\ 
\end{tabular}
\label{tab}
\end{ruledtabular}
\end{table}
\twocolumngrid

\end{document}